\documentclass[prl,aps,groupedaddress,twocolumn,floatfix,nofootinbib]{revtex4}
\usepackage{amsmath,amsthm,amsfonts,amssymb,graphics,graphicx,epsfig,color,times,natbib}

\newcommand{\be}{\begin{equation}}
\newcommand{\ee}{\end{equation}}
\newcommand{\ba}{\begin{eqnarray}}
\newcommand{\ea}{\end{eqnarray}}

\newcommand{\demi}{\frac{1}{2}}

\newcommand{\integers}{\begin{picture}(8,8)\put(0,0){Z}\put(4,0){\line(1,-1){7}}\end{picture}}

\newcommand{\one}{\leavevmode\hbox{\small1\normalsize\kern-.33em1}}

\newcommand{\moy}[1]{\langle #1 \rangle}

\newtheorem{thm}{Theorem}

\newtheorem{lem}[thm]{Lemma}

\newtheorem{prot}{Protocol}
\newtheorem{protocol}[prot]{Protocol}

\begin{document}

\title{Quantifying the nonlocality of GHZ quantum correlations \\ by a bounded communication simulation protocol}


\author{Cyril Branciard$^1$ and Nicolas Gisin$^2$}
\affiliation{$^1$School of Mathematics and Physics, The University of Queensland, St Lucia, QLD 4072, Australia \\
$^2$Group of Applied Physics, University of Geneva, 20 rue de l'Ecole-de-M\'edecine, CH-1211 Geneva 4, Switzerland}

\date{\today}

\begin{abstract}
The simulation of quantum correlations with alternative nonlocal resources, such as classical communication, gives a natural way to quantify their nonlocality. While multipartite nonlocal correlations appear to be useful resources, very little is known on how to simulate multipartite quantum correlations. We present the first known protocol that reproduces 3-partite GHZ correlations with bounded communication: 3 bits in total turn out to be sufficient to simulate all equatorial Von Neumann measurements on the 3-partite GHZ state.
\end{abstract}

\maketitle

When measurements are performed on several quantum systems in an entangled state, the statistics of the results may contain correlations that can't be simulated by shared local variables. Such correlations are called nonlocal. They can be identified by their capacity to violate some inequality: these so-called Bell inequalities are indeed satisfied by all correlations that can be explained by shared local variables~\cite{bell}.

The observation that quantum theory predicts nonlocal correlations is not new; it goes all the way back to the famous EPR argument~\cite{EPR}. Many experimental confirmations have been demonstrated all over the world during the last two decades of last century~\cite{aspect_nature_review}. During the first ten years of this century, the interest for nonlocal correlations has shifted from mere skepticism and incredulity to more constructive questions. First, physicists raised the question of the power of nonlocal correlations for information processing; the main examples being ``device-independent'' quantum key distribution~\cite{Barrett_NSQKD, Acin_DIQKD_PRL, Pironio_DIQKD_NJP} and random number generation~\cite{Pironio_randomness_Nature}. Second, theorists realized that quantum correlations, although possibly nonlocal, are never maximally nonlocal, hence the question ``why is quantum theory not more nonlocal?''~\cite{Popescu_more_nonlocal}; note the great advances since the original question ``Why is quantum theory not local?''.

Thirdly, and this is the topic of this letter, physicists and computer scientists tried to quantify nonlocality; that is, to treat nonlocality as a physical quantity. Indeed, the violation of a Bell inequality only proves that the correlations are not local, but doesn't tell us anything about how far from local they are, i.e.  how much nonlocality they contain. Intuitively, a larger violation should signal more nonlocality. But this na\"ive approach is insufficient as some correlations may violate different Bell inequalities by different amounts. A quite natural measure of nonlocality is the number of classical bits that need to be communicated from one party to another in order to simulate the correlation. For local correlations, no communication is needed, as shared local variables suffice; hence local correlations have a ``communication measure'' equal to zero, as it should be. Let us stress that the idea is not to imagine that nature does use communication to produce nonlocal correlations, it is only to quantify the amount of nonlocality by the quantity of communication required to simulate the correlation.

That such a measure of nonlocality is natural is testified by the fact that it has been introduced by 3 independent papers~\cite{Maudlin,Brassard_Cleve_Tapp_8bits,Steiner}. More precisely, in this letter we adopt as a measure the number of bits communicated between all partners in the worst case~\cite{Maudlin, Brassard_Cleve_Tapp_8bits}. An alternative could be to count the number of bits sent on average~\cite{Steiner,Gisin_Gisin}.

For the case of 2 qubits in a maximally entangled state, Toner and Bacon~\cite{toner_bacon} proved that one single bit of communication suffices (if one restricts the analyses to projective Von Neumann measurements, as we do in this letter). Hence the nonlocality of two spin-$\demi$ particles in the singlet state is 1 bit. For the general case of 2 qubits in a partially entangled state it is known that 2 bits of communication are enough~\cite{toner_bacon}, though it is still unproven that one bit isn't sufficient. At first, one may think that the nonlocality of a partially entangled state shouldn't be larger than that of maximally entangled states, but this is not so clear once one realized the difficulty of simulating at the same time the nonlocal correlation and the nontrivial marginal probabilities~\cite{scarani_review,brunner_partial_entgmt}. 

\paragraph{GHZ correlations.} In this letter we consider 3-qubit GHZ quantum correlations and present the first known protocol to simulate such nonlocal correlations with bounded communication. This problem is the straightforward next step after the 2-qubit case; it attracted the attention of most of the specialists. After years of unsuccessful efforts, the feeling started to spread that it might be impossible with finite communication~\cite{Broadbent_GHZ}. Some hope, however, appeared when Bancal et al.~\cite{bancal_GHZ} presented a protocol with unbounded, but finite average communication. Moreover, a team recently presented a nonconstructive existence proof of a protocol with 6 bits of communication~\cite{Palazuelos_etal}; 
the proof turned out to be flawed, but the impulse was given!

More precisely, our goal is to simulate, with classical communication, the quantum correlations obtained by performing equatorial Von Neumann measurements on a 3-partite GHZ state. Namely: 3 parties, Alice, Bob and Charlie, each receive an input angle $\phi_A, \phi_B$ and $\phi_C \in [0,2\pi]$ (corresponding to a measurement setting on the equator of the Bloch sphere ${\cal S}^2$), and they must output binary outcomes $\alpha, \beta, \gamma \in \{+1, -1\}$, such that the expectation values satisfy
\ba
& \moy{\alpha\beta\gamma} = \cos(\phi_A+\phi_B+\phi_C), \label{eq_GHZ_correl}
\ea
while all single- and bi-partite marginals vanish. Note that although the choice of equatorial measurements is restrictive, these are enough to come up with the ``GHZ paradox''~\cite{GHZ_paradox}. We will show that our problem can be solved with finite communication. For that, we first introduce a protocol that provides ``stronger'' correlations, before showing how to adequately transform these and obtain the desired cosine correlations.

\paragraph{Simulation with classical communication.}
Consider the following protocol, that uses 3 bits of communication: 2 from Bob to Alice, and 1 from Charlie to Alice. The sign function below is defined as ${\mathrm{sign}}(x)=+1$ if $x\geq0$, ${\mathrm{sign}}(x)=-1$ if $x<0$.

\begin{protocol}

Let Alice and Bob share two random vectors $\vec \lambda_1$ and $\vec \lambda_2$, uniformly distributed on the sphere ${\cal S}^2$, together with a random bit $\xi \in \{0,1\}$; let Alice and Charlie share a random variable $\varphi_c$, uniformly distributed on $[0,2\pi]$.

After reception of their measurement settings $\phi_A, \phi_B$ and $\phi_C$, the three parties proceed as follows:

\begin{enumerate}
\setcounter{enumi}{-1}

\item Bob defines $\hat b$ 
to be the equatorial vector with azimuthal angle $\frac{\pi}{2}-2\phi_B$; he calculates $\sigma_0 = {\mathrm{sign}}(\hat b \cdot \vec \lambda_1){\mathrm{sign}}(\hat b \cdot \vec \lambda_2)$, and sends the bit $\tau_0 = \frac{1-\sigma_0}{2}$ to Alice.
Alice and Bob can then both determine the azimuthal angle $\varphi_0 \in [0,2\pi]$ of $\vec \lambda_0 = \vec \lambda_1 + (-1)^{\tau_0} \vec \lambda_2$; they calculate $\varphi_b = \frac{\varphi_0}{2} + \xi \pi \in [0,2\pi]$.

\item Alice, Bob and Charlie define $\tilde{\phi}_A = \phi_A - \varphi_b - \varphi_c, \tilde{\phi}_B = \phi_B + \varphi_b$ and $\tilde{\phi}_C = \phi_C + \varphi_c$, respectively.

\item Bob calculates $\sigma_b = {\mathrm{sign}}(\sin 2\tilde{\phi}_B)$, and sends $\tau_b = \frac{1-\sigma_b}{2}$ to Alice; he outputs $\beta = {\mathrm{sign}}(\sin \tilde{\phi}_B)$.

Similarly, Charlie calculates $\sigma_c = {\mathrm{sign}}(\sin 2\tilde{\phi}_C)$, and sends $\tau_c = \frac{1-\sigma_c}{2}$ to Alice; he outputs $\gamma = {\mathrm{sign}}(\sin \tilde{\phi}_C)$.

\item Alice outputs $\alpha = {\mathrm{sign}}(\sin(-\tilde{\phi}_A - \tau_b\frac{\pi}{2} - \tau_c\frac{\pi}{2}))$.

\end{enumerate}

\end{protocol}

Before analyzing the correlation given by Protocol~1, let us give an intuitive understanding of it. Forget for now the rather technical step~0~\cite{footnote_protocol0}, and note that after step~1, one has $\tilde{\phi}_A + \tilde{\phi}_B + \tilde{\phi}_C = \phi_A + \phi_B + \phi_C$; the first two steps will ensure that the final tripartite correlation depends on the sum $\phi = \phi_A + \phi_B + \phi_C$ only, and that all marginals vanish. Assume now that $\tilde{\phi}_B, \tilde{\phi}_C \in [0,\pi]$ (and hence $\beta = \gamma = +1$); if this is not the case, Bob and Charlie can locally subtract $\pi$ to $\tilde{\phi}_B$ or $\tilde{\phi}_C$ and flip their output, so that the correlation $E(\phi) = \moy{\alpha\beta\gamma}$ is unchanged -- this is precisely why we ask them to output $\beta = {\mathrm{sign}}(\sin \tilde{\phi}_B)$ and $\gamma = {\mathrm{sign}}(\sin \tilde{\phi}_C)$. In step~2, Bob and Charlie tell Alice in which quadrant ($[0,\frac{\pi}{2}]$ or $[\frac{\pi}{2},\pi]$) their angles $\tilde{\phi}_B$ and $\tilde{\phi}_C$ are. From this information, Alice knows in which half-circle $\tilde{\phi}_B+\tilde{\phi}_C$ is (more precisely, she knows ${\mathrm{sign}}(\sin(\tilde{\phi}_B+\tilde{\phi}_C))$ or ${\mathrm{sign}}(\cos(\tilde{\phi}_B+\tilde{\phi}_C))$, depending on whether $\tau_b = \tau_c$ or $\tau_b \neq \tau_c$); if $-\tilde{\phi}_A$ is in the same half-circle, she wants to obtain a good correlation with $\beta\gamma=+1$ (if by chance $-\tilde{\phi}_A=\tilde{\phi}_B+\tilde{\phi}_C$, she wants a perfect correlation!), and will thus output $\alpha=+1$; otherwise, she will output $\alpha=-1$; this corresponds precisely to step~3.

As shown in Appendix~A, Protocol~1 gives vanishing marginals, and the following 3-partite correlation $E_1(\phi) := \moy{\alpha\beta\gamma}$:
\ba
E_1(\phi) &=& \frac{32}{\pi^2} \sum_{n \geq 0} \frac{1}{(2n\!+\!1)^2} \, \frac{1}{4\!-\!(2n\!+\!1)^2} \cos\big((2n\!+\!1)\phi\big) \nonumber \\
&=& 1-\frac{2\phi-\sin2\phi}{\pi} \quad {\mathrm{for}} \ \phi \in [0,\pi]
\ea
which, as already mentioned, only depends on the sum $\phi = \phi_A + \phi_B + \phi_C$.

\begin{figure}
\includegraphics[width=8cm]{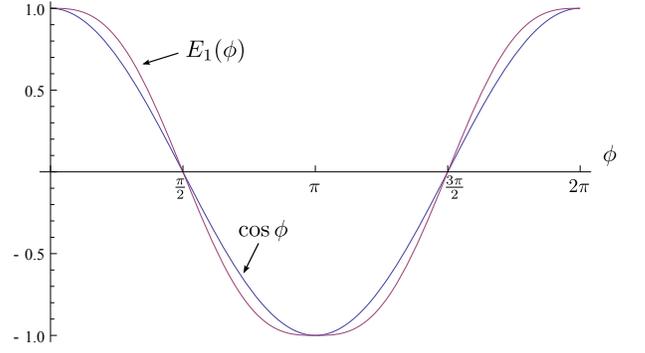}
\caption{Correlation $E_1(\phi) = E_1(\phi_A + \phi_B + \phi_C)$ obtained from Protocol~1, compared to the desired correlation $\cos(\phi)$.}
\label{fig_E1_vs_cos}
\end{figure}

$E_1(\phi)$ is shown on Figure~\ref{fig_E1_vs_cos}. One can notice that it is ``stronger'' than the desired $\cos\phi$ correlation, in the sense that $|E_1(\phi)| \geq |\cos\phi|$ for all $\phi$. Intuitively, one should be able to add some noise and weaken the correlation. However, weakening any given stronger correlation so as to obtain the desired cosine is not trivial, since this weakening must depend on $\phi$ and should in particular not weaken the extreme correlations for $\phi=0$ and $\pi$.
In fact, it seems that correlations must in general have quite specific properties for them to possibly be transformed to the desired cosine.


In order to do so, and starting from a $2\pi$-periodic correlation function such that $E(0) = -E(\pi) = 1$, one can for instance try to mix correlations of the form $E\big((2m+1)\phi\big)$, with $m \in \integers$, as this will preserve the perfect (anti-)correlations for $\phi = 0$ and $\pi$. The following lemma gives a sufficient condition under which such a mixture can indeed give the desired cosine correlation~\cite{footnote_nec_and_suff_conditions}.

\begin{lem} \label{lemma1}
Let $E(\phi)$ be a ($2\pi$-periodic, $\pi$-anti-periodic, even) real function with a Fourier decomposition of the form
\ba
E(\phi) = \sum_{n \geq 0} e_{2n+1} \cos\big((2n+1)\phi\big),
\label{decomp_E}
\ea
such that
\ba \left\{
\begin{array}{l}
e_1 > 0, \ e_{2n+1} \leq 0 \ {\mathrm{for \ all}} \ n \geq 1, \\
E(0) = \sum_{n \geq 0} e_{2n+1} = 1, \\
E''(0) = -\sum_{n \geq 0} (2n+1)^2 e_{2n+1} \leq 0.
\end{array}
\right.
\label{assumptions_lemma1}
\ea

Then $\cos \phi$ can be decomposed as
\ba
\cos \phi = \sum_{m \geq 0} p_{2m+1} E\big((2m+1)\phi\big),
\label{decomp_cos}
\ea
with $p_{2m+1} \geq 0$ for  all $m \geq 0$.

\end{lem}

In particular, for $\phi = 0$, one gets $\sum_{m\geq0} p_{2m+1} = 1$. The coefficients $p_{2m+1} \geq 0$ can thus be interpreted as probabilities, and the kind of ``inverse Fourier decomposition'' (\ref{decomp_cos}) is indeed a probabilistic mixture of correlations $E\big((2m+1)\phi\big)$.

A proof of Lemma~1 is given in Appendix B together with the explicit form of the $p_{2m+1}$. It is easy to check that $E_1$ satisfies the conditions (\ref{assumptions_lemma1}). Hence, there exist coefficients $p_{2m+1} \geq 0$ such that $\sum_{m\geq0} p_{2m+1} = 1$ and
\ba
\sum_{m \geq 0} p_{2m+1} E_1\big((2m+1)\phi\big) = \cos\phi.
\label{decomp_cos_E1}
\ea
Consequently, and since $(2m+1)\phi_A + (2m+1)\phi_B + (2m+1)\phi_C = (2m+1)\phi$, the following protocol gives the desired cosine correlation and solves our problem, with the same 3 classical bits as in Protocol~1:

\begin{protocol}
Let Alice, Bob and Charlie share, in addition to the randomness already introduced in Protocol 1, a random variable $M$ that takes the value $M = 2m+1$ with probability $p_{2m+1}$, where $\{p_{2m+1}\}_{m\geq 0}$ are the coefficients of the decomposition (\ref{decomp_cos_E1}).

After reception of their measurement settings $\phi_A, \phi_B$ and $\phi_C$, the three parties run Protocol 1 with input angles $(2m+1)\phi_A, (2m+1)\phi_B$ and $(2m+1)\phi_C$, respectively.
\end{protocol}

\paragraph{Variants of our communication protocol.}

For convenience, let us from now on consider the equivalent 0/1 bit values corresponding to the 3 parties outputs in Protocol~1 (or~2): $a = \frac{1-\alpha}{2}, b = \frac{1-\beta}{2}$ and $c = \frac{1-\gamma}{2}$; the additions below will be modulo 2. Writing explicitly $a = a_{\tau_0\tau_b\tau_c}$ as a function of the classical communication (the bits $\tau_0,\tau_b,\tau_c$) that Alice receives, one has
\ba
a_{\tau_0 11} = a_{\tau_0 00}+1 \hspace{3mm}{\mathrm{and}}\hspace{3mm} a_{\tau_0 01} = a_{\tau_0 10} \ .
\label{alphaTau}
\ea

One can see that our communication protocol can actually be declined in different forms. In particular, Alice might not need to know the individual values of the bits $\tau_b$ and $\tau_c$, but only their sum $\tau_{bc}=\tau_{b}+\tau_{c}$. Charlie's bit $\tau_c$ could for example be sent to Bob instead; Bob would then send $\tau_{bc}$ to Alice, who would output $a_{\tau_0 \tau_{bc}}'=a_{\tau_0 \tau_{bc} 0}$; in the case when $\tau_b=\tau_c=1$, the `+1' term in (\ref{alphaTau}) can be introduced by Bob instead, who should output $b_{\tau_c}'=b+\tau_b\tau_c$. This thus induces a protocol~1', summarized as
\ba
{\mathrm{Prot. \ 1'}} &:& \left\{ \begin{array}{l}
C \stackrel{\tau_c}{\longrightarrow} B \stackrel{\tau_0,\tau_{bc}}{\longrightarrow} A \\
c'=c \, , \ b_{\tau_c}'=b+\tau_b\tau_c \, , \ a_{\tau_0 \tau_{bc}}'=a_{\tau_0 \tau_{bc} 0} \, ,
\end{array} \right. \nonumber
\ea

With similar considerations, one can come up with many different variants with varied communication patterns, such as, for instance:
\ba
{\mathrm{Prot. \ 1''}} &\!\!\!:\!\!& \!\left\{ \begin{array}{l}
C \stackrel{\tau_c}{\longrightarrow} B \, , \ B \stackrel{\tau_0}{\longrightarrow} A \stackrel{\tau_a}{\longrightarrow} B \\
\!c''\!=c, a_{\tau_0}''\!=a_{\tau_0 00}, b_{\tau_c\tau_a}''\!\!=b\!+\!\tau_a\tau_b\!+\!\tau_a\tau_c\!+\!\tau_b\tau_c \, ,
\end{array} \right. \nonumber \\
{\mathrm{Prot. \ 1'''}} &\!\!\!:\!\!& \!\left\{ \begin{array}{l}
B \stackrel{\tau_b}{\longrightarrow} C \stackrel{\tau_{bc}}{\longrightarrow} A \stackrel{\tau_\alpha}{\longrightarrow} B \\
c_{\tau_b}'''=c\!+\!\tau_b\tau_c, \, a_{\tau_{bc}}'''=a_{0 \tau_{bc} 0}, \, b_{\tau_\alpha}'''=b\!+\!\tau_0\tau_\alpha \, ,
\end{array} \right. \nonumber
\ea
with $\tau_a = a_{\tau_0 00}+a_{\tau_0 10}$ and $\tau_\alpha = a_{0\tau_{bc}0}+a_{1\tau_{bc}0}$.
These variants and the original protocol look different, though they all require 3 bits of communication and lead to the same correlation. 
All of them have severe timing constraints (which is common for communication protocols): there are always some players that can't produce their output before some other partners receive their input and send them some information.

\paragraph{Simulation with PR boxes.}

An interesting alternative to measure nonlocality is to estimate the number of nonlocal PR-boxes~\cite{PRbox} (some kind of ``unit of nonlocality''~\cite{barrett_pironio_PR_unit_of_NL}) required to simulate the correlations. Since the correlations we consider in this letter have no single- nor bi-partite marginals, all the variant communication protocols introduced above can be translated into PR-box based protocols~\cite{barrett_pironio_PR_unit_of_NL}. Indeed, using (\ref{alphaTau}) for the original version of Protocol~1, one can always decompose the sum $a+b+c$ as follows: 
\ba
a_{\tau_0\tau_b\tau_c}+b+c &=& a_{000}+b+c + \tau_0(a_{000}\!+\!a_{100})\nonumber\\
&& \hspace{-2.5cm} + \ \tau_b(a_{000}\!+\!a_{010})+\tau_0\tau_b(a_{000}\!+\!a_{010}\!+\!a_{100}\!+\!a_{110}) + \tau_b\tau_c\nonumber\\
&& \hspace{-2.2cm} + \ \tau_c(a_{000}\!+\!a_{010}) + \tau_0\tau_c(a_{000}\!+\!a_{010}\!+\!a_{100}\!+\!a_{110}) \, .  \label{5PR1GHZ}
\ea
The product terms in (\ref{5PR1GHZ}) can be generated by using nonlocal boxes: 5 PR-boxes can be used for the first 5 products (3 between Alice and Bob, 1 between Bob and Charlie and 1 between Alice and Charlie); the last product can be generated by a 3-partite GHZ-box, which can in turn be constructed from 3 PR-boxes~\cite{barrett_etal_PRA05}. Hence, a total of 8 PR-boxes suffices to simulate the tripartite GHZ correlations.

Interestingly, the variant communication protocols described above all lead to a PR-box based protocol with the same configuration of 8 PR-boxes, all used precisely in the same way. In addition to this invariance, and similarly to quantum correlations, the PR-box based protocol does not suffer from any timing constraint. Hence, it might be a more faithful tool to measure quantum nonlocality (at least, for correlations with vanishing marginals) -- this question is quite general and would require further scrutinies beyond the scope of this letter. Note finally that reciprocally, simulating the PR boxes by communication gives a systematic way to generate different variants of our initial protocol, depending on which way we the communication goes.

\paragraph{Detection loophole.}

Another interesting connection is between our communication protocol and simulation models based on the detection loophole~\cite{Pearle_DetLH,Gisin_Gisin}. For this connection let us start for instance from the last variant of the communication protocols. In the detection-loophole-based protocol, $\tau_b$, $\tau_{bc}$ and $\tau_\alpha$ are 3 additional shared random variables and each player outputs a bit if and only if the appropriate $\tau$ agrees with the bit he should send in the communication protocol. Hence, using variant 1''' of our protocol, the detection-loophole-based protocol simulates the GHZ correlations with ``detection efficiencies'' of 50\% for Alice, Bob and Charlie. 
Other variant protocols can lead to detection-loophole-based protocols with asymmetric detection efficiencies.

\paragraph{Conclusion.}

We have proven that 3 bits of communication (or 8 PR-boxes) suffice to simulate 3-qubit GHZ equatorial correlations; hence the nonlocality of these correlations is at most of 3 bits (8 PR-boxes). In the course of our derivation, we introduced a strategy to obtain a cosine correlation as a mixture of other (``harmonic'') correlations, via Lemma~1, that we believe could be used in other contexts as well.

In this letter we considered correlations with vanishing single- and bi-partite marginals. If one considers also measurements on the GHZ state out of the equatorial plane~\cite{footnote_non_equatorial}, or if one considers other states such as biased GHZ-like states for instance, then the marginals will no longer be random, and simulating the entire probability distribution is likely to be significantly harder~\cite{brunner_partial_entgmt}.

Two other important open problems are the questions of the optimality of our protocol and of its generalization to more parties. For 3 parties, since the GHZ correlations are truly 3-partite~\cite{svetlichny}, a minimum of 2 bits is necessary to connect the 3 parties. We could find a 2-bit protocol (Protocol 1, without step~0, see~\cite{footnote_protocol0}) that gives stronger correlations than $\cos\phi$ and that can approximate it to a very good accuracy, but not perfectly. For the $N$-partite case, it is easy to generalize protocol 1, again without step~0, using $(N-1)\log_2(N-1)$ bits of communication: divide the equator of the Bloch sphere into $2(N-1)$ equal sectors, let each of the $N-1$ last parties share a random angle $\varphi_i$ with Alice, and tell her in which sector their angle $\tilde{\phi_I}=\phi_I+\varphi_i$ (modulo $\pi$) is. This leads again to a protocol giving stronger correlations than $\cos\phi$ (actually, stronger and stronger as $N$ increases), with a number of bits that is asymptotically equivalent to the lower bound derived in~\cite{Broadbent_GHZ} for the simulation of GHZ correlations ($N\log_2 N-2N$). Unfortunately, we did not find a generalization 
that would give a correlation satisfying the assumptions of Lemma~1, so that the exact cosine correlation could then be obtained as in Protocol~2.

These observations lead us to formulate the following question: should we understand a ``stronger'' correlation 
as being ``more non-local''? If our goal is to quantify the power of nonlocality as a resource for achieving some information processing task, then the next question follows: is there any (useful) task, for which a stronger correlation might actually be less powerful than a weaker one? If this is not the case, then one could be happy with simulation protocols that give stronger correlations than the desired ones, and for this operational interpretation of the nonlocality measure, we could conclude that the nonlocality of the 3-partite GHZ correlations is at most 2 bits (or 3 PR-boxes), and that of the $N-$partite GHZ correlations is at most $(N-1)\log_2(N-1)$ bits.

Nonlocal correlations are fascinating. First, because they can't be simulated by mere shared local variables; next, because even if finite communication is allowed, their simulation remains tedious and quite artificial. Hence, simulating in particular quantum nonlocal correlations with classical resources, like shared local variables and communication, looks in general extremely difficult. This underlines the power of nonlocal correlations. Yet, such simulations seem to give a good measure of nonlocality (whether we are interested in the exact simulation or in the ``operational nonlocality'' measure), possibly the best together with PR-box based simulations, and provide the only story that takes place in space and time about how they could occur.

We acknowledge discussions with G. Brassard, M. Kaplan, S. Pironio and I. Villanueva. This work profited from financial support from the Australian Research Council Centre of Excellence for Quantum Computer Technology, the Swiss NCCR-QP and NCCR-QSIT, and the EU AG-QORE.

\bibliography{bib_GHZsimul}

\begin{thebibliography}{29}
\bibitem[{foo({\natexlab{a}})}]{footnote_protocol0}
\bibinfo{note}{Step~0 allows Alice and Bob to sample their shared random
  variable $\varphi_b$~\cite{degorre_PRA05} so that it follows an appropriate
  sine distribution (see Appendix A). Note that we could define a similar
  protocol as Protocol~1, but without step~0, and starting directly with
  $\varphi_b$ uniformly distributed on $[0,2\pi]$. This simpler protocol, that
  requires only 2 bits of communication (or 3 PR-boxes, 1 between each pair of
  parties), would also give a stronger correlation $E_0(\phi) =
  \frac{32}{\pi^3} \sum_{n \geq 0} \frac{(-1)^n}{(2n+1)^3}
  \cos\big((2n+1)\phi\big)$ ($= 1-\frac{4\phi^2}{\pi^2}$ for
  $\phi\in[0,\frac{\pi}{2}]$) than $\cos\phi$. However, $E_0(\phi)$ does not
  satisfy the assumptions of Lemma~\ref{lemma1}; by mixing correlations of the
  form $E_0\big((2m+1)\phi\big)$, one can approximate the cosine correlation
  with a very good accuracy, but not exactly.}
\bibitem[{foo({\natexlab{b}})}]{footnote_nec_and_suff_conditions}
\bibinfo{note}{It is interesting to note the similarities of our approach here
  with that presented in~\cite{regev_toner}, and in particular to compare the
  assumptions of our Lemma~1 with those of \mbox{Lemma~3.1} there. As
  in~\cite{regev_toner}, our lemma only gives sufficient conditions for the
  decomposition (\ref{decomp_cos}) to exist, with $p_{2m+1} \geq 0$; it is not
  clear to us which are the necessary and sufficient conditions for the
  conclusion of Lemma~1 to be reached. In particular, the fact that $E(\phi)$
  is stronger than $\cos\phi$ is actually not a necessary condition (for a
  counterexample, consider for instance $E(\phi) = 1.04 \cos\phi -
  0.04\cos5\phi$).}
\bibitem[{foo({\natexlab{c}})}]{footnote_non_equatorial}
\bibinfo{note}{For non-equatorial measurements, our protocol fails to reproduce
  the non-vanishing marginals. However, if one is not interested in the
  marginals but only wants to simulate the tripartite correlation term
  $\langle\alpha\beta\gamma\rangle =
  \sin\theta_A\sin\theta_B\sin\theta_C\cos\phi$ (where $\theta_A,\theta_B$ and
  $\theta_C$ are the zenith angles of the 3 measurement settings in ${\cal
  S}^2$), then our protocol can be used, as each party can locally add the
  appropriate amount of noise to introduce the factors $\sin\theta_I$. Note
  also that if no more than one party performs a non-equatorial measurement,
  the marginals are in fact still random, and the full correlation can again be
  simulated.}
\end{thebibliography}


\onecolumngrid 

\bigskip
\bigskip
\hrule
\bigskip

\twocolumngrid

\appendix

\section{Appendix A: Calculation of $E_1(\phi)$}

In this Appendix we analyze the correlation obtained with Protocol~1. We first introduce two useful lemmas.
\bigskip

\begin{lem} \label{lemma_app1}
Let $\vec a$ be a vector on the equator of ${\cal S}^2$, with azimuthal angle $\phi_A$. Consider two random vectors $\vec\lambda_1, \vec\lambda_2$ uniformly distributed on the half sphere ${\cal S}^+(\vec a) = \{\vec\lambda \in {\cal S}^2 \ | \ \vec a \cdot \vec\lambda \geq 0\}$, and define $\vec \lambda = \frac{\vec\lambda_1 + \vec\lambda_2}{||\vec\lambda_1 + \vec\lambda_2||}$. Then the azimuthal angle $\varphi \in [\phi_A-\frac{\pi}{2},\phi_A+\frac{\pi}{2}]$ of $\vec\lambda$ is distributed according to
\ba
\rho(\varphi) = \demi \cos(\varphi-\phi_A).
\ea
\end{lem}

\bigskip

Alternatively, if one doesn't want to restrict {\it a priori} $\varphi$ to be in $[\phi_A-\frac{\pi}{2},\phi_A+\frac{\pi}{2}]$, one can write, for any $\varphi \in [0,2\pi]$ (or any real interval with amplitude $2\pi$),
\ba
\rho(\varphi) = \max(0,\demi \cos(\varphi-\phi_A)).
\ea

\bigskip

\emph{Proof of Lemma \ref{lemma_app1}:}

Denote by $\rho_{{\cal S}^2}$ and $\rho_{{\cal S}^+}$ the uniform distributions on ${\cal S}^2$ and ${\cal S}^+(\vec a)$, respectively. For another vector $\vec b$ on the equator of ${\cal S}^2$, with azimuthal angle $\phi_B$, let's calculate:
\ba
&& \hspace{-.4cm} {\cal I} \nonumber \\
&& \hspace{-.5cm} := \! \int_{{\cal S}^+(\vec a)} \! \rho_{{\cal S}^+}(\vec\lambda_1) \,{\mathrm d}\vec\lambda_1 \! \int_{{\cal S}^+(\vec a)} \! \rho_{{\cal S}^+}(\vec\lambda_2) \,{\mathrm d}\vec\lambda_2 \ {\mathrm{sign}} \big(\vec b \! \cdot \! (\vec\lambda_1 \! + \! \vec\lambda_2)\big) \label{eq_int_TB} \nonumber \\
&& \hspace{-.4cm} = \! \int_{{\cal S}^2} \! \rho_{{\cal S}^2}(\vec\lambda_1) \,{\mathrm d}\vec\lambda_1 \! \int_{{\cal S}^2} \! \rho_{{\cal S}^2}(\vec\lambda_2)  \,{\mathrm d}\vec\lambda_2 \ {\mathrm{sign}} \big(\vec b \! \cdot \! (\sigma_1\vec\lambda_1 \! + \! \sigma_2\vec\lambda_2)\big) \nonumber \\
&& \hspace{-.4cm} = \! \int_{{\cal S}^2} \! \rho_{{\cal S}^2}(\vec\lambda_1) \,{\mathrm d}\vec\lambda_1 \! \int_{{\cal S}^2} \! \rho_{{\cal S}^2}(\vec\lambda_2)  \,{\mathrm d}\vec\lambda_2 \ \sigma_1 \, {\mathrm{sign}} \big(\vec b \! \cdot \! (\vec\lambda_1 \! + \! \sigma_1\sigma_2\vec\lambda_2)\big) \nonumber
\ea
with $\sigma_i = {\mathrm{sign}} (\vec a \cdot \vec\lambda_i)$. The last integrals were calculated by Toner and Bacon in~\cite{toner_bacon}, and were found to be equal to $\vec a \cdot \vec b = \cos(\phi_B-\phi_A)$.

Now, the term ${\mathrm{sign}} \big(\vec b \cdot (\vec\lambda_1+\vec\lambda_2)\big)$ in the definition of ${\cal I}$ above 
is simply equal to ${\mathrm{sign}} \big(\!\cos(\varphi-\phi_B)\big)$, so that the integral is also equal to
\ba
{\cal I} &\!\!=\!\!& \cos(\phi_B\!-\!\phi_A) \!=\! \int_{\phi_A-\frac{\pi}{2}}^{\phi_A+\frac{\pi}{2}} \rho(\varphi) \,{\mathrm d}\varphi \ {\mathrm{sign}} \big(\!\cos(\varphi-\phi_B)\big) \, . \nonumber \\ \label{eq_I_rho}
\ea
Defining $R(\phi) = \int_{\phi_A-\frac{\pi}{2}}^{\phi} \rho(\varphi) \,{\mathrm d}\varphi$ (with $R(\phi_A+\frac{\pi}{2}) = 1$ due to normalization), the explicit calculation of (\ref{eq_I_rho}) leads to
\ba
\cos(\phi_B\!-\!\phi_A) \!= 1\!-\!2R(\phi_B-\frac{\pi}{2}) & \ {\mathrm{if}} \ & \phi_A \leq \phi_B \leq \phi_A \!+\! \pi \, , \nonumber \\
\cos(\phi_B\!-\!\phi_A) \!= 2R(\phi_B+\frac{\pi}{2})\!-\!1 & \ {\mathrm{if}} \ & \phi_A \!-\! \pi \leq \phi_B \leq \phi_A \, . \nonumber
\ea
With now $\varphi = \phi_B \mp \frac{\pi}{2}$, both equalities above give
\ba
2R(\varphi)-1 = \sin(\varphi-\phi_A) & \ {\mathrm{for}} \ & \phi_A - \frac{\pi}{2} \leq \varphi \leq \phi_A + \frac{\pi}{2} \, . \nonumber
\ea
After differentiation, one obtains
\ba
\rho(\varphi) = \demi\cos(\varphi-\phi_A). \nonumber
\ea
\hfill$\Box$

\bigskip

\begin{lem} \label{lemma_app2}
After step~0 of Protocol~1, $\varphi_b \in [0,2\pi]$ is distributed according to
\ba
\rho(\varphi_b) = \frac{1}{4}|\sin 2(\varphi_b + \phi_B)| \, .
\label{distr_varphi_b}
\ea
\end{lem}

\bigskip

\emph{Proof of Lemma \ref{lemma_app2}:}

We now use the notation $\sigma_i = {\mathrm{sign}} (\hat b \cdot \vec\lambda_i)$.

Suppose first that $\sigma_1 = +1$. Then $\vec \lambda_0 = \vec \lambda_1 + \sigma_1\sigma_2 \vec \lambda_2 = \sigma_1 \vec \lambda_1 + \sigma_2 \vec \lambda_2$ is the sum of two vectors uniformly distributed on ${\cal S}^+(\hat b)$. From Lemma~\ref{lemma_app1}, the distribution of its azimuthal angle $\varphi_0 \in [0,2\pi]$ is
\ba
\rho_+(\varphi_0) &=& \max\big(0, \demi\cos(\varphi_0 - (\frac{\pi}{2}-2\phi_B))\big) \nonumber \\
&=& \max\big(0, \demi\sin(\varphi_0 + 2\phi_B)\big) \, . \nonumber
\ea
In the case where $\sigma_1 = -1$, $\vec \lambda_0 = \vec \lambda_1 + \sigma_1\sigma_2 \vec \lambda_2 = -\sigma_1 \vec \lambda_1 - \sigma_2 \vec \lambda_2$ is the sum of two vectors uniformly distributed on ${\cal S}^+(-\hat b)$. The distribution of its azimuthal angle $\varphi_0 \in [0,2\pi]$ is
\ba
\rho_-(\varphi_0) = \max\big(0, \demi\sin(\varphi_0 + 2\phi_B + \pi)\big) \, . \nonumber
\ea
As Alice and Bob ignore the individual value of $\sigma_1$ ($=\pm1$ with equal probabilities), the overall distribution of $\varphi_0$ is
\ba
\rho_0(\varphi_0) = \demi \big(\rho_+(\varphi_0)+\rho_-(\varphi_0)\big) = \frac{1}{4}|\sin(\varphi_0 + 2\phi_B)| \, . \nonumber
\ea
The distribution of $\varphi_b^0=\frac{\varphi_0}{2} \in [0,\pi]$ is then
\ba
\rho_b^0(\varphi_b^0) = 2\rho_0(2\varphi_b^0) = \frac{1}{2}|\sin(2\varphi_b^0 + 2\phi_B)| \, , \nonumber
\ea
and after adding $\xi \pi$, with $\xi \in \{0,1\}$ random, the distribution of $\varphi_b=\varphi_b^0 + \xi\pi \in [0,2\pi]$ is finally
\ba
\rho(\varphi_b) = \frac{1}{4}|\sin(2\varphi_b + 2\phi_B)| \, . \nonumber
\ea

\hfill$\Box$

\bigskip

Let us now calculate the correlation obtained with Protocol~1. One can, for simplicity, directly integrate over the variables $\tilde{\phi}_B = \phi_B + \varphi_b$ and $\tilde{\phi}_C = \phi_C + \varphi_c$; from Lemma~\ref{lemma_app2}, $\tilde{\phi}_B$ is distributed according to $\rho(\tilde{\phi}_B) = \frac{1}{4}|\sin(2\tilde{\phi}_B)|$, while $\tilde{\phi}_C$ is uniformly distributed on $[0,2\pi]$. One can easily check that the single- and bi-partite marginals vanish; the tripartite correlation writes
\ba
&& \hspace{-.4cm} \moy{\alpha\beta\gamma} = \nonumber \\
&& \hspace{-.2cm} \int_{0}^{2\pi} \! {\mathrm d}\tilde{\phi}_B \frac{1}{4}|\sin(2\tilde{\phi}_B)| \! \int_{0}^{2\pi} \! \frac{{\mathrm d}\tilde{\phi}_C}{2\pi} \ {\mathrm{sign}}(\sin \tilde{\phi}_B) \ {\mathrm{sign}}(\sin \tilde{\phi}_C) \nonumber \\
&& \hspace{.8cm} \times \ {\mathrm{sign}}\big(\!\sin(\tilde{\phi}_B \!+\! \tilde{\phi}_C \!-\! \phi \!-\! \tau_b(\tilde{\phi}_B)\frac{\pi}{2} \!-\! \tau_c(\tilde{\phi}_C)\frac{\pi}{2})\big) , \nonumber
\ea
and only depends on $\phi = \phi_A + \phi_B + \phi_C$.

Using the periodicity of the integrand function, one obtains
\ba
E_1(\phi) &:=& \moy{\alpha\beta\gamma} = \frac{2}{\pi} \int_{0}^{\frac{\pi}{2}} {\mathrm d}\tilde{\phi}_B \int_{0}^{\frac{\pi}{2}} {\mathrm d}\tilde{\phi}_C \ \sin(2\tilde{\phi}_B) \nonumber \\
&& \hspace{2.7cm} \times \, {\mathrm{sign}}\big(\sin(\tilde{\phi}_B + \tilde{\phi}_C - \phi)\big) . \nonumber
\ea
It is convenient to use the Fourier decomposition ${\mathrm{sign}}(\sin x) = \frac{4}{\pi} \sum_{n\geq0}\frac{1}{2n\!+\!1}\sin\big((2n\!+\!1)x\big)$ to calculate the integrals. One then easily gets
\ba
E_1(\phi) &\!\!=\!& \frac{32}{\pi^2} \sum_{n \geq 0} \frac{1}{(2n\!+\!1)^2} \, \frac{1}{4\!-\!(2n\!+\!1)^2} \cos\big((2n\!+\!1)\phi\big) \, . \nonumber \\
\ea


One can finally check that $E_1(\phi)$ can also be written as $E_1(\phi) = 1-\frac{2\phi-\sin2\phi}{\pi}$ for $\phi \in [0,\pi]$, from which the full function can be obtained by symmetry and periodicity.




\bigskip
\bigskip
\section{Appendix B: Proof of Lemma 1}

Assume that $E(\phi)$ satisfies the assumptions (\ref{decomp_E}-\ref{assumptions_lemma1}). Define $p_1 = \frac{1}{e_1}$ and, for $m \geq 1$,
\ba
p_{2m+1} &\!=\!& -\frac{1}{e_1} \sum_{k = 0}^{m-1}\sum_{n = 1}^{m} p_{2k+1} e_{2n+1} \delta_{(2k+1)(2n+1),2m+1} \nonumber \\
\label{eq_ps}
\ea
(with $\delta_{i,j} = 1$ if $i=j$, $\delta_{i,j} = 0$ otherwise).

Note that these coefficients $p_{2m+1}$ are non-negative. We will show that they lead to the decomposition (\ref{decomp_cos}). The proof is partly inspired by that of Lemma 3.1 in~\cite{regev_toner}; we divide it into 3 steps.

\medskip

\begin{itemize}
\item {\it Step~1:} We first prove that the coefficients $p_{2m+1}$ can be written as 
\end{itemize}
\vspace{-1.2cm}
\ba
& \hspace{-.8cm} p_{2m+1} = \frac{1}{e_1}\sum_{\ell_3, \ell_5, \dots} \!\!\! \frac{(\ell_3 + \ell_5 + \dots)!}{\ell_3! \ell_5! \dots} \! \left(-\frac{e_3}{e_1}\right)^{\ell_3} \!\! \left(-\frac{e_5}{e_1}\right)^{\ell_5} \! \! \dots \, , 
\label{eq_ps_2}
\ea
where the (finite) sum is taken over all non-negative integers $\ell_3, \ell_5, \dots$ such that $3^{\ell_3} \times 5^{\ell_5} \times \dots = 2m+1$.

\medskip

We prove it by induction. Note that (\ref{eq_ps_2}) holds for $m=0$; suppose $p_{2k+1}$ can be written as in (\ref{eq_ps_2}) for all $k < m$. Then, with the notation $i|j$ meaning ``$i$ divides $j$'', one can write

\begin{widetext}
\ba
p_{2m+1} &=& - \, \frac{1}{e_1} \ \sum_{\stackrel{n = 1,}{{\mathrm{s.t.}} \ 2n+1|2m+1}}^{m} \ e_{2n+1} \ p_{\frac{2m+1}{2n+1}} \nonumber \\
&=& \frac{1}{e_1} \ \sum_{\stackrel{n = 1,}{{\mathrm{s.t.}} \ 2n+1|2m+1}}^{m} \ \left(-\frac{e_{2n+1}}{e_1}\right) \sum_{\stackrel{\ell_3, \ell_5, \dots}{{\mathrm{s.t.}} \ 3^{\ell_3}5^{\ell_5}\dots=\frac{2m+1}{2n+1}}} \frac{(\ell_3 + \ell_5 + \dots)!}{\ell_3! \ell_5! \dots} \! \left(-\frac{e_3}{e_1}\right)^{\ell_3} \!\! \left(-\frac{e_5}{e_1}\right)^{\ell_5} \! \dots \nonumber
\ea
\ba
p_{2m+1} &=& \frac{1}{e_1} \ \sum_{\stackrel{n = 1,}{{\mathrm{s.t.}} \ 2n+1|2m+1}}^{m} \ \ \sum_{\stackrel{\ell_3, \dots \ell_{2n+1}, \dots}{{\mathrm{s.t.}} \ 3^{\ell_3}\dots(2n+1)^{\ell_{2n+1}+1}\dots=2m+1}} \hspace{-.5cm} \frac{(\ell_3 + \dots + \ell_{2n+1} + \dots)!}{\ell_3! \dots \ell_{2n+1}! \dots} \left(-\frac{e_3}{e_1}\right)^{\ell_3} \dots \left(-\frac{e_{2n+1}}{e_1}\right)^{\ell_{2n+1}+1} \dots \nonumber \\
&=& \frac{1}{e_1} \ \sum_{\stackrel{n = 1,}{{\mathrm{s.t.}} \ 2n+1|2m+1}}^{m} \ \ \sum_{\stackrel{\ell_3, \ell_5, \dots}{{\mathrm{s.t.}} \ 3^{\ell_3}5^{\ell_5}\dots=2m+1}} \hspace{-.1cm} \frac{\ell_{2n+1}}{\ell_3 + \ell_5 + \dots} \ \frac{(\ell_3 + \ell_5 + \dots)!}{\ell_3! \ell_5! \dots} \left(-\frac{e_3}{e_1}\right)^{\ell_3} \left(-\frac{e_5}{e_1}\right)^{\ell_5} \dots \nonumber
\ea
where we relabeled $(\ell_{2n+1}+1) \to \ell_{2n+1}$.
After exchanging the two sums, and using the fact that $\sum_{n = 1}^{m} \ \frac{\ell_{2n+1}}{\ell_3 + \ell_5 + \dots} = 1$, one obtains (\ref{eq_ps_2}) as desired.
\bigskip
\bigskip
\end{widetext}

\begin{itemize}
\item {\it Step~2:} We now show that $\sum p_{2m+1}$ converges absolutely.
\end{itemize}


Eq. (\ref{eq_ps_2}) can be written as
\ba
& \hspace{-2cm} p_{2m+1} = \frac{1}{e_1} \frac{1}{(2m+1)^{3/2}} \sum_{\ell_3, \ell_5, \dots} \!\!\! \frac{(\ell_3 + \ell_5 + \dots)!}{\ell_3! \ell_5! \dots} \nonumber \\
& \hspace{3.5cm} \times \left(-\frac{3^{3/2} e_3}{e_1}\right)^{\ell_3} \!\! \left(-\frac{5^{3/2} e_5}{e_1}\right)^{\ell_5} \! \dots \, . \nonumber
\ea
All the terms in the sum are non-negative. If one extends the sum to all integers $\ell_3, \ell_5, \dots$, one gets the upper bound
\ba
& \hspace{-2cm} p_{2m+1} \leq \frac{1}{e_1} \frac{1}{(2m+1)^{3/2}} \sum_{L=0}^\infty \ \sum_{\ell_3, \ell_5, \dots} \frac{L!}{\ell_3! \ell_5! \dots} \nonumber \\
& \hspace{4cm} \times \left(-\frac{3^{3/2} e_3}{e_1}\right)^{\ell_3} \!\! \left(-\frac{5^{3/2} e_5}{e_1}\right)^{\ell_5} \! \dots \nonumber
\ea
where in the second sum, the indices are now such that $\ell_3 + \ell_5 + \dots = L$.

From the assumptions (\ref{assumptions_lemma1}), $\sum (2n+1)^{3/2}(-e_{2n+1}) \leq \sum (2n+1)^2(-e_{2n+1})$ converges absolutely, and one can apply the multinomial theorem to calculate the inner sum; one finds that this sum is $(C_{3/2})^L$, where $C_{3/2} = \sum_{n=1}^\infty \big(-\frac{(2n+1)^{3/2}e_{2n+1}}{e_1}\big)$.

Now, $0 \leq \sum_{n=1}^\infty (2n+1)^{3/2}(-e_{2n+1}) < \sum_{n=1}^\infty (2n+1)^2(-e_{2n+1}) \leq e_1$ by assumption\footnote{The strict inequality in what precedes should actually be replaced by an equality in the case where $e_1 = 1$ and $e_{2n+1} = 0$ for all $n\geq 1$. But the conclusion still holds in that trivial case.}; hence $0 \leq C_{3/2} < 1$, and therefore one gets
\ba
& 0 \leq p_{2m+1} \leq \frac{1}{e_1} \frac{1}{(2m+1)^{3/2}} \sum_{L=0}^\infty \ (C_{3/2})^L \nonumber \\
& \hspace{1.5cm} = \frac{1}{e_1} \frac{1}{1-C_{3/2}} \frac{1}{(2m+1)^{3/2}} \, , \nonumber
\ea
which implies that $\sum p_{2m+1}$ converges absolutely.

\medskip

Note also that the assumptions (\ref{assumptions_lemma1}) imply that $\sum e_{2n+1}$ converges absolutely as well.

\bigskip
\bigskip
\bigskip
\bigskip

\begin{itemize}
\item {\it Step~3:} Conclusion
\end{itemize}
The fact that $\sum p_{2m+1}$ and $\sum e_{2n+1}$ are absolutely convergent allows one to calculate the following infinite sums:

\ba
&& \hspace{-.4cm} \sum_{m \geq 0} p_{2m+1} E\big((2m+1)\phi\big) \nonumber \\
&& \hspace{-.4cm} = \sum_{m \geq 0} p_{2m+1} \sum_{n \geq 0} e_{2n+1} \cos\big((2m+1)(2n+1)\phi\big) \qquad \nonumber \\
&& \hspace{-.4cm} = \sum_{k \geq 0} \! \Big( \! \sum_{m = 0}^{k} \! \sum_{n = 0}^{k} p_{2m\!+\!1} e_{2n\!+\!1} \delta_{(2m\!+\!1)(2n\!+\!1),2k\!+\!1} \! \Big) \! \cos\!\big((2k\!+\!1)\phi\big) \, . \nonumber
\ea
By definition (\ref{eq_ps}), the double sum inside the brackets is equal to $\delta_{k,1}$, and one obtains, as desired,
\ba
\sum_{m \geq 0} p_{2m+1} E\big((2m+1)\phi\big) = \cos(\phi) \, . \label{eq_sum_pmEm}
\ea

\hfill$\Box$

\newpage

\end{document}